\documentclass[11pt]{article}
\usepackage{color,epsfig,fullpage,palatino,rotating}
\usepackage{epsfig,graphicx,color}
\usepackage{citesort}
\newtheorem{theorem}{Theorem}
\newtheorem{lemma}{Lemma}
\newcommand{\setof}[1]{\{{#1}\}}
\newcommand{\set}[2]{\{{#1}\mid{#2}\}}
\def\myendproof{{\ \vbox{\hrule\hbox{%
   \vrule height1.3ex\hskip0.8ex\vrule}\hrule }}\par}
\newenvironment{proof}{\noindent{\bf Proof. }}{\myendproof}

\title{An Optimal Algorithm for the Maximum-Density Segment
Problem\thanks{An early version of this paper was presented at 11th
Annual European Symposium on Algorithms, Budapest, Hungary, September
15-20, 2003.}}

\date{\today}

\author{Kai-min Chung\thanks{Institute of Information Science,
Academia Sinica Taipei 115, Taiwan, Republic of China. Part of this
work was done while this author was an undergraduate student at
Department of Computer Science and Information Engineering, National
Taiwan University.}
\and Hsueh-I Lu\thanks{Corresponding author.  
Institute of Information Science, Academia Sinica, Taipei 115, 
Taiwan, Republic of China.  Email:
{\tt{hil@iis.sinica.edu.tw}}. URL: {\tt{www.iis.sinica.edu.tw/\~{
}hil/}}.  Research supported in part by NSC grant 91-2215-E-001-001.}}

\begin{document}
\maketitle

\begin{abstract}
We address a fundamental problem arising from analysis of
biomolecular sequences.  The input consists of two numbers $w_{\min}$
and $w_{\max}$ and a sequence $S$ of $n$ number pairs $(a_i,w_i)$ with
$w_i>0$.  Let {\em segment} $S(i,j)$ of $S$ be the consecutive
subsequence of $S$ between indices $i$ and $j$.  The {\em density} of
$S(i,j)$ is
$d(i,j)=(a_i+a_{i+1}+\cdots+a_j)/(w_i+w_{i+1}+\cdots+w_j)$.  The {\em
maximum-density segment problem} is to find a maximum-density segment
over all segments $S(i,j)$ with $w_{\min}\leq w_i+w_{i+1}+\cdots+w_j
\leq w_{\max}$.  The best previously known algorithm for the problem,
due to Goldwasser, Kao, and Lu, runs in
$O(n\log(w_{\max}-w_{\min}+1))$ time.  In the present paper, we solve
the problem in $O(n)$ time.  Our approach bypasses the complicated
{\em right-skew decomposition}, introduced by Lin, Jiang, and Chao. As
a result, our algorithm has the capability to process the input
sequence in an online manner, which is an important feature for
dealing with genome-scale sequences.  Moreover, for a type of input
sequences $S$ representable in $O(m)$ space, we show how to exploit
the sparsity of $S$ and solve the maximum-density segment problem for
$S$ in $O(m)$ time.
\end{abstract}

\section{Introduction}
\label{sec:intro}
We address the following fundamental problem: The input
consists of two numbers $w_{\min}$ and $w_{\max}$ and a sequence $S$
of number pairs $(a_i,w_i)$ with $w_i>0$ for $i=1,\ldots,n$. A {\em
segment} $S(i,j)$ is a consecutive subsequence of $S$ starting with
index $i$ and ending with index $j$.  For a segment $S(i,j)$, the {\em
width} is $w(i,j)=w_i+w_{i+1}+\cdots+w_j$, and the {\em density} is
$d(i,j)=(a_i+a_{i+1}+\cdots+a_j)/w(i,j)$.  It is not difficult to see
that with an $O(n)$-time preprocessing to compute all $O(n)$ prefix
sums $a_1+a_2+\cdots +a_j$ and $w_1+w_2+\cdots +w_j$, the density of
any segment can be computed in $O(1)$ time.  $S(i,j)$ is {\em
feasible} if $w_{\min}\leq w(i,j) \leq w_{\max}$.  The {\em
maximum-density segment} problem is to find a maximum-density segment
over all $O(n^2)$ feasible segments.

This problem arises from the investigation of non-uniformity of
nucleotide composition within genomic sequences, which was first
revealed through thermal melting and gradient centrifugation
experiments~\cite{Inman:1966:DMP,Macaya:1976:AOE}.  The GC content of
the DNA sequences in all organisms varies from 25\% to 75\%. GC-ratios
have the greatest variations among bacteria's DNA sequences, while the
typical GC-ratios of mammalian genomes stay in 45-50\%.  
Despite intensive research effort in the past two decades, the
underlying causes of the observed heterogeneity remain
debatable~\cite{Bernardi:1986:CCG,Barhardi:2000:IEG,%
Eyre-Walker:1992:EBG,Filipski:1987:CBM,Sueoka:1988:DMP,%
Wolfe:1989:MRD,Francino:1999:IRM,Holmquist:1992:CBT,%
Eyre-Walker:1993:RMG,Charlesworth:1994:GRP}.
Researchers~\cite{Nekrutenko:2000:ACH,Stojanovic:1999:CFM} observed
that the compositional heterogeneity is highly correlated to the GC
content of the genomic sequences.  Other investigations showed that
gene length~\cite{Duret:1995:SAV}, gene
density~\cite{Zoubak:1996:GDH}, patterns of codon
usage~\cite{Sharp:1995:DSE}, distribution of different classes of
repetitive elements~\cite{Soriano:1983:DIR,Duret:1995:SAV}, number of
isochores~\cite{Barhardi:2000:IEG}, lengths of
isochores~\cite{Nekrutenko:2000:ACH}, and recombination rate within
chromosomes~\cite{Fullerton:2001:LRR} are all correlated with GC
content.  More research related to GC-rich segments can be found
in~\cite{Madsen:1997:ICE,Murata:2001:TAF,Ikehara:1996:PON,%
Wang:2002:RFP,Scotto:1993:GRD,Henke:1997:BIP,Wu:1999:IMD,%
Guldberg:1998:DMG,Jin:1997:WIN} and the references therein.

In the most basic form of the maximum-density segment problem, the
sequence $S$ corresponds to the given DNA sequence, where $a_i = 1$ if
the corresponding nucleotide in the DNA sequence is G or C; and $a_i =
0$ otherwise.  In the work of Huang~\cite{Huang:1994:AIR}, sequence
entries took on values of $p$ and $1-p$ for some real number $0 \leq p
\leq 1$.  More generally, we can look for regions where a given set of
patterns occur very often. In such applications, $a_i$ could be the
relative frequency that the corresponding DNA character appears in the
given patterns.  Further natural applications of this problem can be
designed for sophisticated sequence analysis such as mismatch
density~\cite{Sellers:1984:PRG}, ungapped local
alignments~\cite{Alexandrov:1998:SSU}, annotated multiple sequence
alignments~\cite{Stojanovic:1999:CFM}, promoter
mapping~\cite{IoshikhesZ00}, and promoter
recognition~\cite{OhlerNLR01}.

\begin{figure}[t]
\centerline{\input{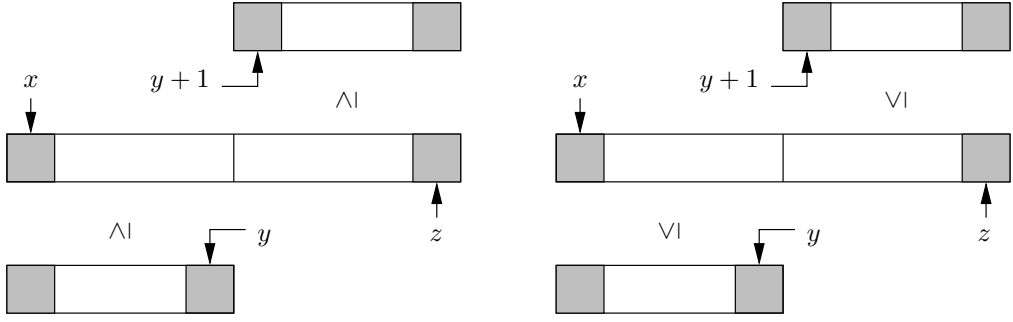}}
\caption{An illustration for Equation~(\ref{eq1}): There are only two
possibilities for the order among $d(x,y),d(x,z),d(y+1,z)$.}
\label{figure:fig1}
\end{figure}

For the {\em uniform} case, i.e., $w_i=1$ for all indices $i$,
Nekrutendo and Li~\cite{Nekrutenko:2000:ACH}, and Rice, Longden and
Bleasby~\cite{Rice:2000:EEM} employed algorithms for the case
$w_{\min}=w_{\max}$, which is trivially solvable in $O(n)$ time. More
generally, when $w_{\min} \neq w_{\max}$, the problem is also easily
solvable in $O(n(w_{\max}-w_{\min}+1))$ time, linear in the number of
feasible segments.  Huang~\cite{Huang:1994:AIR} studied the case where
$w_{\max} = n$, i.e., there is effectively no upper bound on the width
of the desired maximum-density segments. He observed that an optimal
segment exists with width at most $2w_{\min}-1$.  Therefore, this case
is equivalent to the case with $w_{\max}=2w_{\min}-1$ and can be
solved in $O(nw_{\min})$ time in a straightforward manner. Lin, Jiang,
and Chao~\cite{Lin:2002:EAL} gave an $O(n\log w_{\min})$-time
algorithm for this case based on right-skew decompositions of a
sequence.  (See~\cite{LinHJC03} for a related software.)  The case
with general $w_{\max}$ was first investigated by Goldwasser, Kao, and
Lu~\cite{GoldwasserKL02}, who gave an $O(n)$-time algorithm for the
uniform case. Recently, Kim~\cite{Kim03} showed an alternative
algorithm based upon a geometric interpretation of the problem, which
basically relates the maximum-density segment problem to the
fundamental {\em slope selection problem} in computational
geometry~\cite{BronnimannC98,Matousek91,KatzS93,ColeSSS89}. Unfortunately,
Kim's analysis of time complexity has some flaw which seems difficult to
fix.\footnote{Kim claims that all the progressive updates of the lower
convex hulls $L_j\cup R_j$ can be done in overall linear time. The
paper only sketches how to obtain $L_{j+1}\cup R_{j+1}$ from $L_j\cup
R_j$. (See the fourth-to-last paragraph of page 340 in~\cite{Kim03}.)
Unfortunately, Kim seems to overlook the marginal cases when the upper
bound $w_{\max}$ forces the $p_z$ of $L_j\cup R_j$ to be deleted from
$L_{j+1}\cup R_{j+1}$. As a result, obtaining $L_{j+1}\cup R_{j+1}$
from $L_j\cup R_j$ could be much more complicated than Kim's sketch. A
naive implementation of Kim's algorithm still takes
$\Omega(n(w_{\max}-w_{\min}+1))$ time in the worst case.  We believe
that any correct implementation of Kim's algorithm requires
$\Omega(n\log (w_{\max}-w_{\min}+1))$ time in the worse case.}  For
the general (i.e., {\em non-uniform}) case, Goldwasser, Kao, and
Lu~\cite{GoldwasserKL02} also gave an
$O(n\log(w_{\max}-w_{\min}+1))$-time algorithm.  By bypassing the
complicated preprocessing step required in~\cite{GoldwasserKL02}, we
successfully reduce the required time for the general case down to
$O(n)$.  Our result is based upon the following set of equations,
stating that the order of $d(x,y)$, $d(y+1,z)$, and $d(x,z)$ with
$x\leq y <z$ can be determined by that of any two of them:
\begin{equation}
\label{eq1}
\begin{array}{c}
d(x,y) \leq d(y+1,z) \Leftrightarrow d(x,y) \leq d(x,z)
\Leftrightarrow d(x,z) \leq d(y+1,z);\\
d(x,y) \geq d(y+1,z) \Leftrightarrow d(x,y) \geq d(x,z)
\Leftrightarrow d(x,z) \geq d(y+1,z).
\end{array}
\end{equation}
(Both equations can be easily verified by observing the existence of
some number $\rho$ with $0<\rho<1$ and $d(x,z)=\rho\cdot d(x,y)+
(1-\rho)\cdot d(y+1,z)$. See Figure~\ref{figure:fig1}.)  Our algorithm
is capable of processing the input sequence in an online manner, which
is an important feature for dealing with genome-scale sequences.

For bioinformatics applications, e.g.,
in~\cite{Sellers:1984:PRG,Alexandrov:1998:SSU,Stojanovic:1999:CFM,%
IoshikhesZ00,OhlerNLR01}, the input sequence $S$ is usually very {\em
sparse}. 
That is, $S$ can be represented by $m=o(n)$ triples
$(a'_1,w'_1,n_1), (a'_2,w'_2,n_2),\ldots, (a'_m,w'_m,n_m)$ with
$0=n_0< n_1<n_2<\cdots<n_m=n$ to signify that
$(a_i,w_i)=(a'_j,w'_j)$ holds for all indices $i$ and $j$ with
$n_{j-1} < i\leq n_j$ and $1\leq j\leq m$.  If $w'_j=1$ holds for all
$1\leq j\leq m$, we show how to exploit the sparsity of $S$ and solve
the maximum-density problem for $S$ given in the above compact
representation in $O(m)$ time.

The remainder of the paper is organized as follows.
Section~\ref{sec:main-algorithm} shows the main algorithm.
Section~\ref{section:ineffective} explains how to cope with the simple
case that the width upper bound $w_{\max}$ is ineffective.
Section~\ref{section:general} takes care of the more complicated case
that $w_{\max}$ is effective.  Section~\ref{sec:line-version-compact}
explains how to exploit the sparsity of the input sequence for the
uniform case.

\section{The main algorithm}
\label{sec:main-algorithm}

\begin{figure}[t]
\hrule
\begin{tabbing}
x\=xxx\=xxx\=xxx\=xxx\=xxx\=xxx\=\kill
\\
\>\>{\bf algorithm} $\mbox{\sc main}$\\
\>1\>\>let $i_{j_0-1} = 1$;\\
\>2\>\>{\bf for} $j = j_0$ {\bf to} $n$ {\bf do} \{\\
\>3\>\>\>let $i_j = \mbox{\sc best}(\max(i_{j-1}, \ell_j), r_j, j)$;\\
\>4\>\>\>{\bf output} $(i_j, j)$;\\
\>5\>\>\}\\
\\
\>\>{{\bf function} $\mbox{\sc best}(\ell, r, j)$}\\
\>1\>\>let $i=\ell$;\\
\>2\>\>{\bf while} $i < r$ {\bf and} $d(i,\phi(i,r-1))\leq d(i,j)$ {\bf do} \\ 
\>3\>\>\>let $i=\phi(i,r-1)+1$;\\
\>4\>\>{\bf return} $i$;
\end{tabbing}
\hrule
\caption{Our main algorithm.}
\label{figure:fig2}
\end{figure}

\begin{figure}
  \centerline{\input{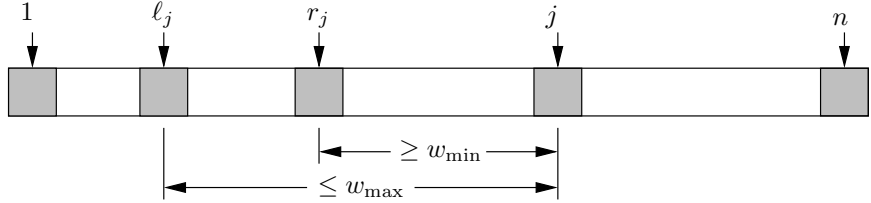}}
\caption{An illustration for the definitions of $\ell_j$ and $r_j$.}
\label{figure:fig3}
\end{figure}

For any integers $x$ and $y$, let $[x,y]$ denote the set
$\setof{x,x+1,\ldots,y}$.  Throughout the paper, we need the following
definitions and notation with respect to the input length-$n$ sequence
$S$ and width bounds $w_{\min}$ and $w_{\max}$.  Let $j_0$ be the
smallest index with $w(1,j_0)\geq w_{\min}$.  Let $J=[j_0,n]$.  For
each $j\in J$, let $\ell_j$ (respectively, $r_j$) be the smallest
(respectively, largest) index $i$ with $w_{\min}\leq w(i,j)\leq
w_{\max}$.  That is, $S(i,j)$ is feasible if and only if
$i\in[\ell_j,r_j]$.  (Figure~\ref{figure:fig3} is an illustration for
the definitions of $\ell_j$ and $r_j$.)  Clearly, for the uniform
case, we have $\ell_{i+1}=\ell_i+1$ and $r_{i+1}=r_i+1$.  As for the
general case, we only know that $\ell_j$ and $r_j$ are both (not
necessarily monotonically) increasing.  One can easily compute all
$\ell_j$ and $r_j$ in $O(n)$ time.  Let $i^*_j$ be the largest index
$k\in[\ell_j,r_j]$ with $d(k,j)=\max\set{d(i,j)}{i\in[\ell_j,r_j]}$.
Clearly, there must be an index $j^*$ such that $S(i^*_{j^*},j^*)$ is
a maximum-density segment of $S$.  Therefore, a natural but seemingly
difficult possibility to solve the maximum-density segment problem
would be to compute $i^*_j$ for all indices $j\in J$ in $O(n)$ time.
Instead, our strategy is to compute an index $i_j\in[\ell_j,r_j]$ for
each index $j\in J$ by the algorithm shown in
Figure~\ref{figure:fig2}, where $\phi(x,y)$ is defined to be the
largest index $z\in[x,y]$ that minimizes $d(x,z)$.
That is, $S(x,\phi(x,y))$ is the longest
minimum-density prefix of $S(x,y)$.  The rest of the section ensures
the correctness of our algorithm by showing $i_{j^*}=i^*_{j^*}$, and
thus reduces the maximum-density segment problem to implementing our
algorithm to run in $O(n)$ time.

\begin{lemma}
\label{lemma:best}
The index returned by function call $\mbox{\sc best}(\ell, r, j)$ is
the largest index $i\in [\ell, r]$ that maximizes $d(i,j)$.
\end{lemma}
\begin{proof}
Let $i^*$ be the largest index in $[\ell, r]$ that maximizes $d(i,j)$,
i.e., $d(i^*, j)=\max_{i\in[\ell,r]}d(i,j)$. Let $i_j$ be the index
returned by function call $\mbox{\sc best}(\ell,r,j)$. We show
$i_j=i^*$ as follows.  If $i_j<i^*$, then $i_j<r$. By the condition of
the while-loop at Step~2 of $\mbox{\sc best}$, we know
$d(i_j,\phi(i_j,r-1))>d(i_j,j)$.  By $d(i_j,j)\leq d(i^*,j)$ and
Equation~(\ref{eq1}), we have $d(i_j,i^*-1)\leq d(i_j,j)$. It follows
that $d(i_j,i^*-1)<d(i_j,\phi(i_j,r-1))$, contradicting the definition of
$\phi(i_j,r-1)$.

On the other hand, suppose that $i_j>i^*$. By definition of $\mbox{\sc
best}$, there must be an index $i\in[\ell,r]$ with $i<r$,
$d(i,\phi(i,r-1))\leq d(i,j)$, and $i\leq i^*<\phi(i,r-1)+1$.  If
$i=i^*$, by Equation~(\ref{eq1}) we have $d(i^*,\phi(i^*,r-1))\leq
d(i^*,j)\leq d(\phi(i^*,r-1)+1,j)$, where the last inequality
contradicts the definition of $i^*$. Now that $i<i^*$, we have
$d(i^*,j)\geq d(i,j)\geq d(i,i^*-1)\geq d(i,\phi(i,r-1))\geq
d(i^*,\phi(i,r-1))$, where (a) the first inequality is by definition
of $i^*$, (b) the second inequality is by Equation~(\ref{eq1}) and the
first inequality, (c) the third inequality is by $i^*\leq\phi(i,r-1)$
and definition of $\phi(i,r-1)$, and (d) the last inequality is by
Equation~(\ref{eq1}) and the third inequality. It follows from
$d(i^*,j)\geq d(i^*,\phi(i,r-1))$ and Equation~(\ref{eq1}) that
$d(\phi(i,r-1)+1,j)\geq d(i^*,j)$, contradicting the definition of
$i^*$ by $i^*<\phi(i,r-1)+1$.
\end{proof}

\begin{theorem}
\label{theorem:correct}
Algorithm $\mbox{\sc main}$ correctly solves the maximum-density
problem.
\end{theorem}
\begin{proof}
We prove the theorem by showing $i_{j^*}=i^*_{j^*}$. Clearly, by
$\ell_{j_0}=i_{j_0-1}=1$ and Lemma~\ref{lemma:best}, the equality
holds if ${j^*}=j_0$. The rest of the proof assumes ${j^*}>j_0$. By
Lemma~\ref{lemma:best} and $\ell_{j^*}\leq i^*_{j^*}$, it suffices to
ensure $i_{j^*-1}\leq i^*_{j^*}$. Assume for contradiction that
there is an index $j\in [j_0,{j^*}-1]$ with $i_{j-1}\leq
i^*_{j^*}<i_j$. By $j<{j^*}$, we know $\ell_j\leq i^*_{j^*}$.  By
Lemma~\ref{lemma:best} and $\max(\ell_j,i_{j-1})\leq i^*_{j^*}<i_j\leq
r_j$, we have $d(i_j,j)\geq d(i^*_{j^*},j)$.  It follows from
Equation~(\ref{eq1}) and $i^*_{j^*}<i_j$ that $d(i^*_{j^*},j)\geq
d(i^*_{j^*},i_j-1)$. By $\ell_{j^*}\leq i^*_{j^*}<i_j\leq r_{j^*}$ and
definition of ${j^*}$, we know $d(i^*_{j^*},{j^*})>d(i_j,{j^*})$. It
follows from $i^*_{j^*}<i_j$ and Equation~(\ref{eq1}) that
$d(i^*_{j^*},i_j-1)>d(i^*_{j^*},{j^*})$.  Therefore, $d(i_j,j)\geq
d(i^*_{j^*},j)\geq d(i^*_{j^*},i_j-1)>d(i^*_{j^*},{j^*})$,
contradicting the definition of ${j^*}$.
\end{proof}

One can verify that the value of $i$ increases by at least one each
time Step~3 of $\mbox{\sc best}$ is executed.  Therefore, to implement
the algorithm to run in $O(n)$ time, it suffices to maintain a data
structure to support $O(1)$-time query for each $\phi(i,r_j-1)$ in
Step 2 of $\mbox{\sc best}$.

\section{Coping with ineffective width upper bound}
\label{section:ineffective}
\begin{figure}[t]
\centerline{\input{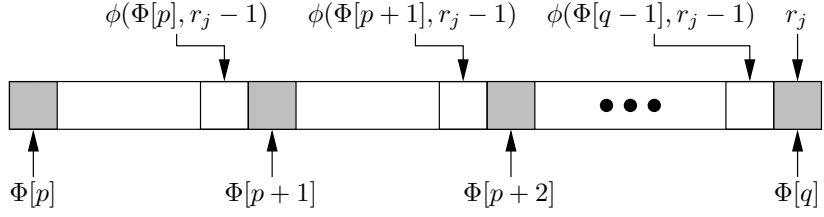}}
\caption{An illustration for Condition~$C_j$.}
\label{figure:fig4}
\end{figure}

When $w_{\max}$ is ineffective, i.e., $w_{\max}\geq w(1,n)$, we have
$\ell_j=1$ for all $j\in F$.  Therefore, the function call in Step~3
of $\mbox{\sc main}$ is exactly $\mbox{\sc
best}(i_{j-1},r_j,j)$. Moreover, during the execution of the function
call $\mbox{\sc best}(i_{j-1}, r_j, j)$, the value of $i$ can only be
$i_{j-1}$, $\phi(i_{j-1},r_j-1)+1$,
$\phi(\phi(i_{j-1},r_j-1)+1,r_j-1)+1, \ldots, r_j$.  Suppose that a
subroutine call to $\mbox{\sc update}(j)$ yields an array $\Phi$ of
indices and two indices $p$ and $q$ of $\Phi$ with $p\leq q$ and
$\Phi[p]=i_{j-1}$ such that the following condition
holds.
\begin{description}
\item[\rm \em Condition~$C_j:$] 
$\Phi[q]=r_j$ and 
$\Phi[t]=\phi(\Phi[t-1],r_j-1)+1$ holds for each index $t\in [p+1,q]$.
\end{description}
(See Figure~\ref{figure:fig4} for an illustration.)  Then, the
subroutine call to $\mbox{\sc best}(i_{j-1}, r_j, j)$ can clearly be
replaced by $\mbox{\sc lbest}(j)$, as defined in
Figure~\ref{figure:fig5}.  That is, $\mbox{\sc lbest}(j)$ can access
the value of each $\phi(i,r_j-1)$ by looking up $\Phi$ in $O(1)$ time.
It remains to show how to implement $\mbox{\sc update}(j)$ such that
all $O(n)$ subroutine calls to $\mbox{\sc update}$ from Step~3 of
$\mbox{\sc lmain}$ run in overall $O(n)$ time.  The following lemma is
crucial in ensuring the correctness and efficiency of our
implementation shown in Figure~\ref{figure:fig5}, where
Condition~$C_{j_0-1}$ stands for $p=1$, $q=0$, and $i_{j_0-1}=1$.

\begin{figure}[t]
\hrule
\begin{tabbing}
x\=xxx\=xxx\=xxx\=xxx\=xxx\=xxx\=\kill
\\
\>\>{\bf algorithm} $\mbox{\sc lmain}$\\
\>1\>\>let $p=1$, $q=0$, and $i_{j_0-1}=1$;\\
\>2\>\>{\bf for} $j = j_0$ {\bf to} $n$ {\bf do} \{\\
\>3\>\>\>{\bf call} $\mbox{\sc update}(j)$;\\
\>4\>\>\>let $i_j = \mbox{\sc lbest}(j)$;\\
\>5\>\>\>{\bf output} $(i_j, j)$;\\
\>5\>\>\}\\
\\
\>\>{\bf function} $\mbox{\sc lbest}(j)$\\
\>1\>\>{\bf while} $p<q$ {\bf and} $d(\Phi[p],\Phi[p+1]-1)\leq d(\Phi[p],j)$ {\bf do}\\
\>2\>\>\>let $p=p+1$;\\
\>3\>\>{\bf return} $\Phi[p]$;\\
\\
\>\>{\bf subroutine} $\mbox{\sc update}(j)$\\
\>1\>\>{\bf for} $r=r_{j-1}+1$ {\bf to} $r_j$ {\bf do} \{\\
\>2\>\>\>{\bf while} $p<q$ {\bf and} $d(\Phi[q-1],\Phi[q]-1)\geq d(\Phi[q-1],r-1)$ {\bf do}\\
\>3\>\>\>\>let $q=q-1$;\\
\>4\>\>\>let $q=q+1$;\\
\>5\>\>\>let $\Phi[q]=r$;\\
\>6\>\>\}
\end{tabbing}
\hrule
\caption{An efficient implementation for the case that $w_{\max}$ is
ineffective.}
\label{figure:fig5}
\end{figure}

\begin{lemma}
\label{lemma:new}
For each index $j\in J$, the following statements hold.
\begin{enumerate}
\item\label{s1}
If Condition~$C_{j-1}$ holds right before calling $\mbox{\sc
update}(j)$, then Condition~$C_j$ holds right after the subroutine
call.
\item\label{s2}
If Condition~$C_j$ holds right before calling $\mbox{\sc lbest}(j)$,
then the index returned by the function call is 
exactly that returned by $\mbox{\sc best}(\Phi[p],\Phi[q],j)$.
\end{enumerate}
\end{lemma}

\begin{proof}
Statement~\ref{s1}.  It is not difficult to verify that with the
initialization $p=1$, $q=0$, and $i_{j_0-1}=1$, Condition $C_{j_0}$
holds with $p=1$ and $q\geq 1$ after calling $\mbox{\sc update}(j_0)$.
The rest of the proof assumes $j\in J-\setof{j_0}$.  For each $r\in
[r_{j-1}+1,r_j]$, one can see from the definition of $\phi$ that
$\phi(\ell,r-1)$ is either $\phi(\ell,r-2)$ or $r-1$. More precisely, we
have $\phi(\ell,r-1)=r-1$ if and only if $d(\ell,\phi(\ell,r-2))\geq
d(\ell,r-1)$. Furthermore, if $\phi(\ell,r-2)< r-2$, then one can prove as
follows that $\phi(\phi(\ell,r-2)+1,r-1)=\phi(\phi(\ell,r-2)+1,r-2)$ implies
$\phi(\ell,r-1)=\phi(\ell,r-2)$.
\begin{quote}
Let $m=\phi(\ell,r-2)$.  By $\phi(m+1,r-1)=\phi(m+1,r-2)$, we have
$d(m+1,\phi(m+1,r-1))<d(m+1,r-1)$. By definition of $\phi$ and
Equation~(\ref{eq1}), we have
$d(\ell,m)<d(m,\phi(m+1,r-1))<d(m+1,\phi(m+1,r-1))$. As a result, we have
$d(\ell,m)<d(m+1,r-1)$, which by Equation~(\ref{eq1}) implies
$d(\ell,m)<d(\ell,r-1)$. Thus $\phi(\ell,r-1)=\phi(\ell,r-2)$.
\end{quote}
Therefore, at the end of each iteration of
the for-loop of $\mbox{\sc update}(j)$, we have that
$\Phi[q]=r$ and $\Phi[t]=\phi(\Phi[t-1],r-1)+1$
holds for each index $t\in [p+1,q]$. (The value of $q$ may change,
though.) It follows that at the end of the for-loop, Condition~$C_j$
holds.

Statement~\ref{s2}.  By Condition~$C_j$, one can easily verify that
$\mbox{\sc lbest}(j)$ is a faithful implementation of
$\mbox{best}(\Phi[p],\Phi[q],j)$.  Therefore, the statement holds.
\end{proof}

\begin{lemma}
\label{lemma:update}
The implementation $\mbox{\sc lmain}$ solves the maximum-density
problem for the case with ineffective $w_{\max}$ in $O(n)$ time.
\end{lemma}
\begin{proof}
By Lemma~\ref{lemma:new}(\ref{s1}) and definitions of $\mbox{\sc
update}$ and $\mbox{\sc lbest}$, both Condition~$C_j$ and
$\Phi[p]=i_{j-1}$ hold right after the subroutine call to $\mbox{\sc
update}(j)$. By Condition~$C_j$ and Lemma~\ref{lemma:new}(\ref{s2}),
$\mbox{\sc lbest}(j)$ is a faithful implementation of $\mbox{\sc
best}(\Phi[p],\Phi[q],j)$. Therefore, the correctness of $\mbox{\sc
lmain}$ follows from $\Phi[p]=i_{j-1}$, $\Phi[q]=r_j$, and
Theorem~\ref{theorem:correct}.

As for the efficiency of $\mbox{\sc lmain}$, observe that $q-p\geq -1$
holds throughout the execution of $\mbox{\sc lmain}$.  Note that each
iteration of the while-loops of $\mbox{\sc lbest}$ and $\mbox{\sc
update}$ decreases the value of $q-p$ by one.  Clearly, Step~4 of
$\mbox{\sc update}$ is the only place that increases the value of
$q-p$.  Since it increases the value of $q-p$ by one for $O(n)$ times,
the overall running time of $\mbox{\sc lmain}$ is $O(n)$.
\end{proof}

\section{Coping with effective width upper bound}
\label{section:general}

In contrast to the previous simple case, when $w_{\max}$ is arbitrary,
$\ell_j$ may not always be $1$.  Therefore, the first argument of the
function call in Step~3 of $\mbox{\sc main}$ could be $\ell_j$ with
$\ell_j>i_{j-1}$.  It seems quite difficult to update the
corresponding data structure $\Phi$ in overall linear time such that
both $\Phi[p]=\max(i_{j-1},\ell_j)$ and Condition~$C_j$ hold
throughout the execution of our algorithm.  To overcome the
difficulty, our algorithm sticks with Condition~$C_j$ but allows
$\Phi[p]>\max(i_{j-1},\ell_j)$.  As a result, $\max_{j\in
J}d(i_j,j)$ may be less than $\max_{j\in J}d(i^*_j, j)$.  Fortunately,
this potential problem can be resolved if we simultaneously solve a
series of variant versions of the maximum-density segment problem. 

\subsection{A variant version of the maximum-density segment problem}
\label{subsection:variant}

\begin{figure}
\centerline{\input{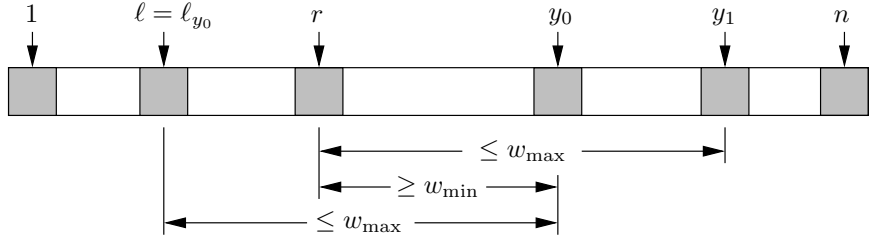}}
\caption{An illustration for the relation among $\ell,r,y_0,y_1$.}
\label{figure:fig6}
\end{figure}

Suppose that we are given two indices $r$ and $y_0$ with $w(r,y_0)
\geq w_{\min}$.  Let $X=[\ell,r]$ and $Y=[y_0,y_1]$ be two intervals
such that $\ell=\ell_{y_0}$ and and $y_1$ is the largest index in $J$
with $w(r, y_1)\leq w_{\max}$.  See Figure~\ref{figure:fig6} for an
illustration.  The {\em variant version} of the maximum-density
segment problem is to look for a maximum-density segment over all
feasible segments $S(x,y)$ with $x\in X$, $y\in Y$, and $w_{\min}\leq
w(x,y)\leq w_{\max}$ such that $d(x,y)$ is maximized.

For each $y\in Y$, let $x^*_y$ be the largest index $x \in X$ with
$w_{\min}\leq w(x,y)\leq w_{\max}$ that maximizes $d(x,y)$.  Let $y^*$
be an index in $Y$ with $d(x^*_{y^*},y^*)=\max_{y\in Y}d(x^*_y,y)$.
Although solving the variant version can naturally be reduced to
computing the index $x^*_y$ for each index $y\in Y$, the required
running time is more than what we can afford.  Instead, we compute an
index $x_y\in X$ with $w_{\min}\leq w(x_y,y)\leq w_{\max}$ for each
index $y\in Y$ such that $x_{y^*}=x^*_{y^*}$.  By $w(r,y_0)\geq
w_{\min}$ and $w(r,y_1)\leq w_{\max}$, one can easily see that, for
each $y\in Y$, $r$ is always the largest index $x \in X$ with
$w_{\min}\leq w(x,y)\leq w_{\max}$. Our algorithm for solving the
variant problem is as shown in Figure~\ref{figure:fig7}, presented in
a way to emphasize the analogy between $\mbox{\sc vmain}$ and
$\mbox{\sc main}$.  For example, the index $x_y$ in $\mbox{\sc vmain}$
is the counterpart of the index $i_j$ in $\mbox{\sc main}$. Also, the
index $r$ in $\mbox{\sc vmain}$ plays the role of the index $r_j$ in
$\mbox{\sc main}$.  We have the following lemma whose proof is very
similar to that of Theorem~\ref{theorem:correct}.

\begin{lemma}
\label{lemma:vcorrect}
Algorithm $\mbox{\sc vmain}$ correctly solves the variant version
of the maximum-density problem.
\end{lemma}

\begin{proof}
We prove the theorem by showing
$x_{y^*}=x^*_{y^*}$. Clearly, by $\ell_{y_0}=x_{y_0-1}=\ell$ and
Lemma~\ref{lemma:best}, the equality holds if ${y^*}=y_0$. The rest of
the proof assumes ${y^*}>y_0$. By Lemma~\ref{lemma:best} and
$\ell_{y^*}\leq x^*_{y^*}$, it suffices to ensure $x_{{y^*}-1}\leq
x^*_{y^*}$. Assume for contradiction that there is an index $y\in
[y_0,{y^*}-1]$ with $x_{y-1}\leq x^*_{y^*}<x_y$. By $y<{y^*}$, we know
$\ell_y\leq x^*_{y^*}$.  By Lemma~\ref{lemma:best} and
$\max(\ell_y,x_{y-1})\leq x^*_{y^*}<x_y\leq r$, we have $d(x_y,y)\geq
d(x^*_{y^*},y)$.  It follows from Equation~(\ref{eq1}) and
$x^*_{y^*}<x_y$ that $d(x^*_{y^*},y)\geq d(x^*_{y^*},x_y-1)$. By
$\ell_{y^*}\leq x^*_{y^*}<x_y\leq r$ and definition of ${y^*}$, we
know $d(x^*_{y^*},{y^*})>d(x_y,{y^*})$. It follows from
$x^*_{y^*}<x_y$ and Equation~(\ref{eq1}) that
$d(x^*_{y^*},x_y-1)>d(x^*_{y^*},{y^*})$.  Therefore, $d(x_y,y)\geq
d(x^*_{y^*},y)\geq d(x^*_{y^*},x_y-1)>d(x^*_{y^*},{y^*})$,
contradicting the definition of ${y^*}$.
\end{proof}

\begin{figure}[t]
\hrule
\begin{tabbing}
x\=xxx\=xxx\=xxx\=xxx\=xxx\=xxx\=\kill
\\
\>\>{\bf algorithm} $\mbox{\sc vmain}(r,y_0)$\\
\>1\>\>let $\ell$ be the smallest index in $[1,n]$ with $w(\ell,y_0)\leq w_{\max}$;\\
\>2\>\>let $y_1$ be the largest index in $[1,n]$ with $w(r,y_1)\leq w_{\max}$;\\
\>3\>\>let $x_{y_0-1} = \ell$;\\
\>4\>\>{\bf for} $y = y_0$ {\bf to} $y_1$ {\bf do} \{\\
\>5\>\>\>let $x_y = \mbox{\sc best}(\max(x_{y-1},\ell_y), r, y)$;\\
\>6\>\>\>{\bf output} $(x_y, y)$;\\
\>7\>\>\}
\end{tabbing}
\hrule
\caption{Our algorithm for the variant version of the maximum-density
segment problem, where function $\mbox{\sc best}$ is as defined in
Figure~\ref{figure:fig2}.}
\label{figure:fig7}
\end{figure}

Again, the challenge lies in supporting each query to $\phi(i,r-1)$ of
$\mbox{\sc best}$ in $O(1)$ time during the execution of $\mbox{\sc
vmain}$. Fortunately, unlike during the execution of $\mbox{\sc main}$
where both parameters of $\phi(i,r-1)$ may change, the second
parameter $r-1$ is now fixed.  Therefore, to support each query to
$\phi(i,r-1)$ in $O(1)$ time, we can actually afford $O(r-\ell+1)$
time to compute a data structure $\Psi$ such that
$\Psi[i]=\phi(i,r-1)$ for each $i\in [\ell,r-1]$.  As a result, the
function $\mbox{\sc best}$ can be implemented as the function
$\mbox{\sc vbest}$ shown in Figure~\ref{figure:fig8}.  The following
lemma ensures the correctness and efficiency of our implementation
$\mbox{\sc variant}$ shown in Figure~\ref{figure:fig8}.

\begin{figure}[t]
\hrule
\begin{tabbing}
x\=xxx\=xxx\=xxx\=xxx\=xxx\=xxx\=\kill
\\
\>\>{\bf algorithm} $\mbox{\sc variant}(r,y_0)$\\
\>1\>\>let $\ell$ be the smallest index in $[1,n]$ with $w(\ell,y_0)\leq w_{\max}$;\\
\>2\>\>let $y_1$ be the largest index in $[1,n]$ with $w(r,y_1)\leq w_{\max}$;\\
\>3\>\>{\bf call} $\mbox{\sc init}(\ell,r-1)$;\\
\>4\>\>let $x_{y_0-1} = \ell$;\\
\>5\>\>{\bf for} $y = y_0$ {\bf to} $y_1$ {\bf do} \{\\
\>6\>\>\>let $x_y = \mbox{\sc vbest}(\max(x_{y-1},\ell_y),r,y)$;\\
\>7\>\>\>{\bf output} $(x_y, y)$;\\
\>8\>\>\}\\
\\
\>\>{\bf function} $\mbox{\sc vbest}(\ell, r, y)$\\
\>1\>\>let $x=\ell$;\\
\>2\>\>{\bf while} $x < r$ {\bf and} $d(x,\Psi[x])\leq d(x,y)$ {\bf do} \\ 
\>3\>\>\>let $x=\Psi[x]+1$;\\
\>4\>\>{\bf return} $x$;\\
\\
\>\>{\bf subroutine} $\mbox{\sc init}(\ell,r)$\\
\>1\>\>let $\Psi[r]=r$;\\
\>2\>\>{\bf for} $s=r-1$ {\bf downto} $\ell$ {\bf do} \{\\
\>3\>\>\>let $t=s$;\\
\>4\>\>\>{\bf while} $t<r$ {\bf and} $d(s,t)\geq d(s,\Psi[t+1])$ {\bf do}\\
\>5\>\>\>\>let $t=\Psi[t+1]$;\\
\>6\>\>\>let $\Psi[s]=t$;\\
\>7\>\>\}
\end{tabbing}
\hrule
\caption{An efficient implementation for the algorithm $\mbox{\sc vmain}$.}
\label{figure:fig8}
\end{figure}

\begin{lemma}
\label{lemma:variant-time}
The implementation $\mbox{\sc variant}$ correctly solves the variant
version of the maximum-density segment problem in
$O(r-\ell+y_1-y_0+1)$ time.
\end{lemma}
\begin{proof}
Clearly, if $\Psi[i]=\phi(i,r-1)$ holds for each index $i\in
[\ell,r-1]$, then $\mbox{\sc vbest}$ is a faithful implementation of
$\mbox{\sc best}$. By Lemma~\ref{lemma:vcorrect}, the correctness of
$\mbox{\sc variant}$ can be ensured by showing that after calling
$\mbox{\sc init}(\ell,r)$, $\Psi[i]=\phi(i,r)$ holds for each index
$i\in [\ell,r]$. By Step~1 of $\mbox{\sc init}$, we have
$\Psi[r]=r=\phi(r,r)$. Now suppose that $\Psi[i]=\phi(i,r)$ holds for
each index $i\in [x+1,r]$ right before $\mbox{\sc init}$ is about to
execute the iteration for index $x\in [\ell,r]$. It suffices to show
$\Psi[x]=\phi(x,r)$ after the iteration.  Let
$Z_x=\setof{x, \Psi[x+1], \Psi[\Psi[x+1]+1],
\Psi[\Psi[\Psi[x+1]+1]+1], \ldots, r}$. Let $|Z_x|$ denote the
cardinality of $Z_x$.  We first show $\phi(x,r)\in Z_x$ as follows.
\begin{quote}
Assume for contradiction that $\phi(x,r)\not\in Z_x$, i.e., there is an
index $z \in Z_x$ with $z<\phi(x,r)<\Psi[z+1]=\phi(z+1,r)$. By
definition of $\phi$ and Equation~(\ref{eq1}), we have
$d(z+1,\phi(x,r))>d(z+1,\phi(z+1,r))>d(\phi(x,r)+1,\phi(z+1,r))$ and
$d(\phi(x,r)+1,\phi(z+1,r))\geq d(x,\phi(z+1,r))\geq d(x,\phi(x,r))$.
By $d(z+1,\phi(x,r))>d(x,\phi(x,r))$ and Equation~(\ref{eq1}), we have
$d(x,\phi(x,r)) > d(x,z)$, contradicting the definition of
$\phi(x,r)$.
\end{quote}
For any index $z\in Z_x$ with $z<\phi(x,r)$, we know $z<r$ and
$\phi(z+1,r)=\Psi[z+1]\leq \phi(x,r)$. By $\phi(z+1,r)\leq
\phi(x,r)\leq r$ and definition of $\phi(z+1,r)$, we have
$d(z+1,\phi(x,r))\geq d(z+1,\phi(z+1,r))$. By definition of
$\phi(x,r)$ and Equation~(\ref{eq1}), we have $d(x,z)\geq
d(x,\phi(x,r))\geq d(z+1,\phi(x,r))$. By $d(x,z)\geq
d(z+1,\phi(z+1,r))$ and Equation~(\ref{eq1}), we have $d(x,z)\geq
d(x,\phi(z+1,r))$.  Therefore, if $z<\phi(x,r)$, then Step~5 of
$\mbox{\sc init}$ will be executed to increase the value of $z$.
Observe that $\phi(x,r)=z<r$ and $\Psi[z+1]>z$ imply
$d(x,z)<d(x,\Phi[z+1])$.  It follows that as soon as $z=\phi(x,r)$
holds, whether $\phi(x,r)=r$ or not, the value of $\Psi[x]$ will
immediately be set to $z$ at Step~6 of $\mbox{\sc init}$.

One can see that the running time is indeed $O(r-\ell+y_1-y_0+1)$ by
verifying that throughout the execution of the implementation, (a) the
while-loop of $\mbox{\sc vbest}$ runs for $O(r-\ell+y_1-y_0+1)$
iterations, and (b) the while-loop of $\mbox{\sc init}$ runs for
$O(r-\ell+1)=O(r-\ell+1)$ iterations.  To see statement~(a), just
observe that the value of index $x$ (i) never decreases, (ii) stays in
$[\ell,r]$, and (iii) increases by at least one each time Step~3 of
$\mbox{\sc vbest}$ is executed.  As for statement~(b), consider the
iteration with index $s$ of the for-loop of $\mbox{\sc init}$. Note
that if Step~6 of $\mbox{\sc init}$ executes $t_s$ times in this
iteration, then $|Z_s|=|Z_{s+1}|-t_s+1$. Since $|Z_s|\geq 1$ holds for
each $s\in X$, we have $\sum_{s\in X}t_s=O(r-\ell+1)$, and thus
statement~(b) holds.
\end{proof}

\subsection{Our algorithm for the general case}

With the help of $\mbox{\sc variant}$, we have a linear-time algorithm
for solving the original maximum-density segment problem as shown in
Figure~\ref{figure:fig9}. Algorithm $\mbox{\sc general}$ is
obtained by inserting four lines of codes (i.e., Steps~4--7 of
$\mbox{\sc general}$) between Steps~3 and~4 of $\mbox{\sc lmain}$ in
order to handle the case $i_{j-1}<\ell_j$. Specifically, when
$i_{j-1}<\ell_j$, we cannot afford to appropriately update the data
structure~$\Phi$.  Therefore, instead of moving $i$ to $\ell_j$,
Steps~4 and~5 move $i$ to $\Phi[p]$, where $p$ is the smallest index
with $\ell_j\leq \Phi[p]$.  Of course, these two steps may cause our
algorithm to overlook the possibility of $i_j\in [i_{j-1},\Phi[p]-1]$,
as illustrated in Figure~\ref{figure:fig10}. This is when the variant
version comes in: As shown in the next theorem, we
can remedy the problem by calling $\mbox{\sc variant}(\Phi[p],j)$.

\begin{figure}[t]
\hrule
\begin{tabbing}
x\=xxx\=xxx\=xxx\=xxx\=xxx\=xxx\=\kill
\\
\>\>{\bf algorithm} $\mbox{\sc general}$\\
\>1\>\>let $p=1$, $q=0$, and $i_{j_0-1} = 1$;\\
\>2\>\>{\bf for} $j = j_0$ {\bf to} $n$ {\bf do} \{\\
\>3\>\>\>{\bf call} $\mbox{\sc update}(j)$;\\
\>4\>\>\>{\bf while} $\Phi[p]<\ell_j$ {\bf do}\\
\>5\>\>\>\>let $p=p+1$;\\
\>6\>\>\>{\bf if} $i_{j-1}<\Phi[p]$ {\bf then}\\
\>7\>\>\>\>{\bf call} $\mbox{\sc variant}(\Phi[p], j)$;\\
\>8\>\>\>let $i_j = \mbox{\sc lbest}(j)$;\\
\>9\>\>\>{\bf output} $(i_j, j)$;\\
\>10\>\>\}
\end{tabbing}
\hrule
\caption{Our algorithm for the general case, where $\mbox{\sc update}$
and $\mbox{\sc lbest}$ are defined in Figure~\ref{figure:fig5} and
$\mbox{\sc variant}$ is defined in Figure~\ref{figure:fig8}.}
\label{figure:fig9}
\end{figure}

\begin{figure}[t]
\begin{center}
\input{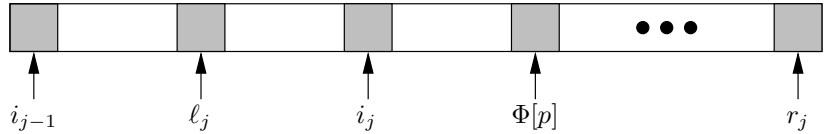}
\end{center}
\caption{An illustration for the situation when Steps~6 and~7 of $\mbox{\sc
general}$ are needed.}
\label{figure:fig10}
\end{figure}

\begin{figure}[t]
\begin{center}
\input{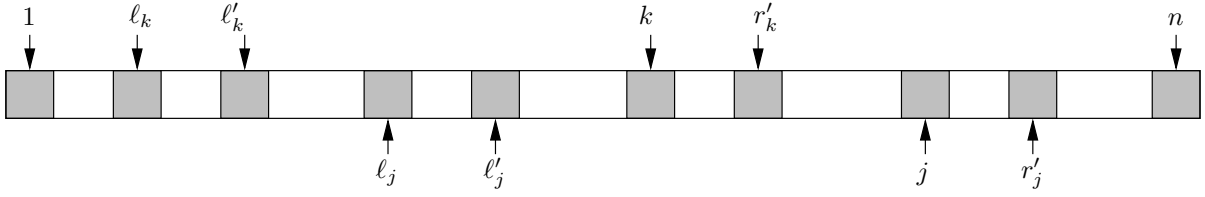}
\end{center}
\caption{An illustration for showing that the overall running time of
all subroutine calls to $\mbox{\sc variant}(\ell'_j,j)$ in
$\mbox{\sc general}$ is $O(n)$.}
\label{figure:fig11}
\end{figure}

\begin{theorem}
\label{theorem:1}
The linear-time algorithm $\mbox{\sc general}$ solves the
maximum-density segment problem in an online manner.
\end{theorem}
\begin{proof}
We prove the correctness of $\mbox{\sc general}$ by showing that
$i^*_{j^*}\ne i_{j^*}$ implies $i^*_{j^*} = x_{j^*}$.  By
Lemma~\ref{lemma:new}(\ref{s1}), after the subroutine call~$\mbox{\sc
update}(j)$ at Step~3 of $\mbox{\sc general}$, Condition~$C_j$ holds.
Clearly, Steps~4 and~5 of $\mbox{\sc general}$, which may increase the
value of $p$, do not affect the validity of Condition~$C_j$. Clearly,
Steps~6 and~7 do not modify $p$, $q$, and $\Phi$. Let $\ell'_j$ be the
value of $\Phi[p]$ right before executing Step~8 of $\mbox{\sc
general}$.  By Lemma~\ref{lemma:new}(\ref{s2}), the index $i_j$
returned by $\mbox{\sc lbest}(j)$ is the largest index in $[\ell'_j,
r_j]$ that maximizes $d(i_j,j)$. Clearly, $i_{j^*}=i^*_{j^*}$ implies
the correctness of $\mbox{\sc general}$.  If $i_{j^*}\ne i^*_{j^*}$,
then there must be an index $j\in [j_0,{j^*}]$ such that
$i_{j-1}\leq i^*_{j^*}<i_{j}$. It can be proved as follows that
$i^*_{j^*}<\ell'_{j}$.
\begin{quote}
Assume $\ell'_{j}\leq i^*_{j^*}$ for contradiction. It follows from
Lemma~\ref{lemma:new}(\ref{s2}) and Equation~(\ref{eq1}) that
$d(i_{j},j)\geq d(i^*_{j^*},j)\geq d(i^*_{j^*},i_{j}-1)$. By
definition of ${j^*}$, we have $d(i^*_{j^*},{j^*})>d(i_{j},j)$,
which by Equation~(\ref{eq1}) implies
$d(i^*_{j^*},i_{j}-1)>d(i^*_{j^*},{j^*})$. Therefore,
$d(i_{j},j)>d(i^*_{j^*},{j^*})$, contradicting the definition of
${j^*}$.
\end{quote}
Since $i_{j-1}\leq i^*_{j^*}<\ell'_{j}$, we know
$w(\ell'_{j}-1,{j^*})\leq w(i^*_{j^*},{j^*})\leq w_{\max}$.  It
follows from Lemma~\ref{lemma:variant-time} that there is an index
pair $(x,y)$ with $w_{\min}\leq w(x,y)\leq w_{\max}$ and
$d(x,y)=d(i^*_{j^*},{j^*})$.

As for the running time, observe that $q-p\geq -1$ holds throughout
the execution of $\mbox{\sc general}$.  Note that each iteration of
the while-loops of $\mbox{\sc general}$, $\mbox{\sc lbest}$ and
$\mbox{\sc update}$ decreases the value of $q-p$ by one.  Clearly,
Step~4 of $\mbox{\sc update}$ is the only place that increases the
value of $q-p$. Moreover, it increases the value of $q-p$ by one for
$O(n)$ times. Therefore, to show that the overall running time of
$\mbox{\sc general}$ is $O(n)$, it remains to ensure that all those
subroutine calls to $\mbox{\sc variant}$ at Step~7 of $\mbox{\sc
general}$ take overall $O(n)$ time. Suppose that $j$ and $k$ are two
arbitrary indices with $k<j$ such that $\mbox{\sc general}$ makes
subroutine calls to $\mbox{\sc variant}(\ell'_k,k)$ and $\mbox{\sc
variant}(\ell'_j,j)$. Let $r'_k$ be the largest index in $[1,n]$
with $w(\ell'_k,r'_k)\leq w_{\max}$. By
Lemma~\ref{lemma:variant-time}, it suffices to show that $\ell'_k <
\ell_j$ and $r'_k<j$ as follows. (See Figure~\ref{figure:fig11}.)  By
definition of $\mbox{\sc general}$, we know that $i_{j-1}<\ell_j$,
which is ensured by the situation illustrated in
Figure~\ref{figure:fig10}. By $k<j$, we have $\ell'_k\leq i_{j-1}$,
implying $\ell'_k<\ell_j$. Moreover, by definitions of $\ell_j$ and
$r'_k$, one can easily verify that $\ell'_k<\ell_j$ implies $r'_k<j$.

It is clear that our algorithm shown in Figure~\ref{figure:fig9} is
already capable of processing the input sequence in an online manner,
since the only preprocessing required is to obtain $\ell_j$, $r_j$,
and the prefix sums of $a_1,a_2,\ldots,a_j$ and $w_1,w_2,\ldots,w_j$
(for the purpose of evaluating the density of any segment in $O(1)$
time), which can easily be computed on the fly.
\end{proof}

\section{Exploiting sparsity for the uniform case}
\label{sec:line-version-compact}
In this section, we assume that $S$ is represented by $m$ pairs
$(a'_1,n_1), (a'_2,n_2),\ldots, (a'_m,n_m)$ with $0=n_0<
n_1<n_2<\cdots<n_m=n$ to signify that $w_1=w_2=\cdots=w_n=1$ and
$a_i=a'_j$ holds for all indices $i$ and $j$ with $n_{j-1} < i\leq
n_j$ and $1\leq j\leq m$. Our algorithm for solving the
maximum-density problem for the $O(m)$-space representable sequence
$S$ is shown in Figure~\ref{figure:fig12}.

\begin{figure}[t]
\hrule
\begin{tabbing}
x\=xxx\=xxx\=xxx\=xxx\=xxx\=xxx\=\kill
\\
\>\>{\bf algorithm} $\mbox{\sc sparse}$\\
\>1\>\>{\bf for} $k=1$ {\bf to} $m$ {\bf do}\\
\>2\>\>\>let $n'_k=n_k-n_{k-1}$;\\
\>3\>\>let $S'$ be the length-$m$ sequence $(n'_1a'_1, n'_1), (n'_2a'_2,
n'_2), \ldots, (n'_ma'_m, n'_m)$;\\
\>4\>\>{\bf output} $(n_{i'-1}+1,n_{j'})$, where $(i',j')$ is an
optimal output of $\mbox{\sc general}(w_{\min}, w_{\max},S')$;\\
\>5\>\>{\bf for} $k=1$ {\bf to} $m$ {\bf do} \{\\
\>6\>\>\>{\bf if} $n_k\geq w_{\min}$ {\bf then}\\
\>7\>\>\>\>{\bf output} $(\ell_{n_k},n_k)$ and $(r_{n_k},n_k)$;\\
\>8\>\>\>{\bf if} $n_{k-1}+w_{\min}\leq n$ {\bf then} \\
\>9\>\>\>\>{\bf output} $(n_{k-1}+1,n_{k-1}+w_{\min})$ and $(n_{k-1}+1,\min(n, n_{k-1}+w_{\max}))$;\\
\>11\>\>\}
\end{tabbing}
\hrule
\caption{Our algorithm that handles sparse input sequence for the
uniform case, where $\mbox{\sc general}$ is defined in
Figure~\ref{figure:fig9}.}
\label{figure:fig12}
\end{figure}

\begin{theorem}
\label{theorem:sparse}
Algorithm $\mbox{\sc sparse}$ solves the maximum-density problem for
the above $O(m)$-space representable sequence in $O(m)$ time.
\end{theorem}
\begin{proof}
By Theorem~\ref{theorem:1}, $\mbox{\sc sparse}$ runs in $O(m)$ time.
Let $S(i^*,j^*)$ be a feasible segment with maximum density.  We first
show that without loss of generality
$i^*-1\in\setof{n_0,n_1,\ldots,n_{m-1}}$ or
$j^*\in\setof{n_1,n_2,\ldots,n_m}$ holds.  More specifically, we show
that if $i^*-1\not\in\setof{n_0,n_1,\ldots,n_{m-1}}$ and
$j^*\not\in\setof{n_1,n_2,\ldots,n_m}$, then $S(i^*+1,j^*+1)$ is also
a feasible segment with maximum density: By
$i^*-1\not\in\setof{n_0,n_1,\ldots,n_{m-1}}$, we know
$a_{i^*-1}=a_{i^*}$.  By $j^*\not\in\setof{n_1,n_2,\ldots,n_m}$, we
know $a_{j^*}=a_{j^*+1}$.  By the optimality of $S(i^*,j^*)$, we have
$a_{i^*}\geq a_{j^*+1}$ and $a_{i^*-1}\leq a_{j^*}$, implying
$a_{i^*-1}=a_{i^*}=a_{j^*}=a_{j^*+1}$. Therefore, $S(i^*+1,j^*+1)$ is
also a maximum-density segment.


\begin{itemize}
\item 
Case 1: $i^*-1\in\setof{n_0,n_1,\ldots,n_{m-1}}$ and
$j^*\in\setof{n_1,n_2,\ldots,n_m}$.  Clearly, Steps~1--4 of $\mbox{\sc
sparse}$ take care of this case.

\item 
Case 2: $i^*-1\not\in\setof{n_0,n_1,\ldots,n_{m-1}}$ and
$j^*\in\setof{n_1,n_2,\ldots,n_m}$.  By Equation~(\ref{eq1}), we know
that $a_{i^*-1}=a_{i^*}\ne d(i^*,j^*)$ implies
$d(i^*-1,j^*)>d(i^*,j^*)$ or $d(i^*+1,j^*)>d(i^*,j^*)$. 
If $(i^*,j^*)$
is not discovered by Steps~6 and~7 of $\mbox{\sc sparse}$, then
$\ell_{j^*}<i^*<r_{j^*}$.  
Since
$i^*-1\not\in\setof{n_0,n_1,\ldots,n_{m-1}}$ implies
$a_{i^*-1}=a_{i^*}$, we know that $\ell_{j^*}<i^*<r_{j^*}$ implies
$d(i^*-1,j^*)=d(i^*,j^*)=d(i^*+1,j^*)$.  Thus, $S(i^*-1,j^*)$ is also
a feasible segment with maximum density. Clearly, we can continue the
same argument until having a maximum-density segment $S(i, j^*)$ such
that either $i-1\in\setof{n_0,n_1,\ldots,n_{m-1}}$, which is handled
in Case~1, or $i=\ell_{j^*}$, which is handled by Steps~6 and~7 of
$\mbox{\sc sparse}$.

\item 
Case 3: $i^*-1\in\setof{n_0,n_1,\ldots,n_{m-1}}$ and
$j^*\not\in\setof{n_1,n_2,\ldots,n_m}$.  By Equation~(\ref{eq1}), we
know that $a_{j^*}=a_{j^*+1}\ne d(i^*,j^*)$ implies
$d(i^*,j^*-1)>d(i^*,j^*)$ or $d(i^*,j^*+1)>d(i^*,j^*)$.  If
$(i^*,j^*)$ is not discovered by Steps~8 and~9 of $\mbox{\sc
sparse}$, then $\ell_{j^*}<i^*<r_{j^*}$.  Since
$j^*\not\in\setof{n_1,n_2,\ldots,n_m}$ implies $a_{j^*}=a_{j^*+1}$, we
know that $\ell_{j^*}<i^*<r_{j^*}$ implies
$d(i^*,j^*-1)=d(i^*,j^*)=d(i^*,j^*+1)$.  Thus, $S(i^*,j^*+1)$ is also
a feasible segment with maximum density. Clearly, we can continue the
same argument until having a maximum-density segment $S(i^*, j)$ such
that either $j\in\setof{n_1,n_2,\ldots,n_m}$, which is handled in
Case~1, or $i^*=\ell_{j}$, which is handled by Steps~8 and~9 of
$\mbox{\sc sparse}$.
\end{itemize}
The theorem is proved.
\end{proof}

\section*{Acknowledgments}
We thank Yi-Hsuan Hsin and Hsu-Cheng Tsai for discussions in the
preliminary stage of this research.  We thank Tien-Ching Lin for
useful comments.

\bibliographystyle{abbrv}
\bibliography{new,density}

\begin{thebibliography}{10}

\bibitem{Alexandrov:1998:SSU}
N.~N. Alexandrov and V.~V. Solovyev.
\newblock Statistical significance of ungapped sequence alignments.
\newblock In {\em Proceedings of Pacific Symposium on Biocomputing}, volume~3,
  pages 461--470, 1998.

\bibitem{Barhardi:2000:IEG}
G.~Barhardi.
\newblock Isochores and the evolutionary genomics of vertebrates.
\newblock {\em Gene}, 241:3--17, 2000.

\bibitem{Bernardi:1986:CCG}
G.~Bernardi and G.~Bernardi.
\newblock Compositional constraints and genome evolution.
\newblock {\em Journal of Molecular Evolution}, 24:1--11, 1986.

\bibitem{BronnimannC98}
H.~Br{\"o}nnimann and B.~Chazelle.
\newblock Optimal slope selection via cuttings.
\newblock {\em Computational Geometry --- Theory and Applications},
  10(1):23--29, 1998.

\bibitem{Charlesworth:1994:GRP}
B.~Charlesworth.
\newblock Genetic recombination: patterns in the genome.
\newblock {\em Current Biology}, 4:182--184, 1994.

\bibitem{ColeSSS89}
R.~Cole, J.~S. Salowe, W.~L. Steiger, and E.~Szemeredi.
\newblock An optimal-time algorithm for slope selection.
\newblock {\em SIAM Journal on Computing}, 18(4):792--810, 1989.

\bibitem{Duret:1995:SAV}
L.~Duret, D.~Mouchiroud, and C.~Gautier.
\newblock Statistical analysis of vertebrate sequences reveals that long genes
  are scarce in {GC}-rich isochores.
\newblock {\em Journal of Molecular Evolution}, 40:308--371, 1995.

\bibitem{Eyre-Walker:1992:EBG}
A.~Eyre-Walker.
\newblock Evidence that both {G+C} rich and {G+C} poor isochores are replicated
  early and late in the cell cycle.
\newblock {\em Nucleic Acids Research}, 20:1497--1501, 1992.

\bibitem{Eyre-Walker:1993:RMG}
A.~Eyre-Walker.
\newblock Recombination and mammalian genome evolution.
\newblock {\em Proceedings of the Royal Society of London Series B, Biological
  Science}, 252:237--243, 1993.

\bibitem{Filipski:1987:CBM}
J.~Filipski.
\newblock Correlation between molecular clock ticking, codon usage fidelity of
  {DNA} repair, chromosome banding and chromatin compactness in germline cells.
\newblock {\em FEBS Letters}, 217:184--186, 1987.

\bibitem{Francino:1999:IRM}
M.~P. Francino and H.~Ochman.
\newblock Isochores result from mutation not selection.
\newblock {\em Nature}, 400:30--31, 1999.

\bibitem{Fullerton:2001:LRR}
S.~M. Fullerton, A.~B. Carvalho, and A.~G. Clark.
\newblock Local rates of recombination are positively correlated with {GC}
  content in the human genome.
\newblock {\em Molecular Biology and Evolution}, 18(6):1139--1142, 2001.

\bibitem{GoldwasserKL02}
M.~H. Goldwasser, M.-Y. Kao, and H.-I. Lu.
\newblock Fast algorithms for finding maximum-density segments of a sequence
  with applications to bioinformatics.
\newblock In R.~Guig{\'o} and D.~Gusfield, editors, {\em Proceedings of the
  Second International Workshop of Algorithms in Bioinformatics}, Lecture Notes
  in Computer Science 2452, pages 157--171, Rome, Italy, 2002. Springer-Verlag.

\bibitem{Guldberg:1998:DMG}
P.~Guldberg, K.~Gronbak, A.~Aggerholm, A.~Platz, P.~{thor Straten},
  V.~Ahrenkiel, P.~Hokland, and J.~Zeuthen.
\newblock Detection of mutations in {GC}-rich {DNA} by bisulphite denaturing
  gradient gel electrophoresis.
\newblock {\em Nucleic Acids Research}, 26(6):1548--1549, 1998.

\bibitem{Henke:1997:BIP}
W.~Henke, K.~Herdel, K.~Jung, D.~Schnorr, and S.~A. Loening.
\newblock Betaine improves the {PCR} amplification of {GC}-rich {DNA}
  sequences.
\newblock {\em Nucleic Acids Research}, 25(19):3957--3958, 1997.

\bibitem{Holmquist:1992:CBT}
G.~P. Holmquist.
\newblock Chromosome bands, their chromatin flavors, and their functional
  features.
\newblock {\em American Journal of Human Genetics}, 51:17--37, 1992.

\bibitem{Huang:1994:AIR}
X.~Huang.
\newblock An algorithm for identifying regions of a {DNA} sequence that satisfy
  a content requirement.
\newblock {\em Computer Applications in the Biosciences}, 10(3):219--225, 1994.

\bibitem{Ikehara:1996:PON}
K.~Ikehara, F.~Amada, S.~Yoshida, Y.~Mikata, and A.~Tanaka.
\newblock A possible origin of newly-born bacterial genes: significance of
  {GC}-rich nonstop frame on antisense strand.
\newblock {\em Nucleic Acids Research}, 24(21):4249--4255, 1996.

\bibitem{Inman:1966:DMP}
R.~B. Inman.
\newblock A denaturation map of the 1 phage {DNA} molecule determined by
  electron microscopy.
\newblock {\em Journal of Molecular Biology}, 18:464--476, 1966.

\bibitem{IoshikhesZ00}
I.~P. Ioshikhes and M.~Q. Zhang.
\newblock Large-scale human promoter mapping using {CpG} islands.
\newblock {\em Nature Genetics}, 26:61--63, 2000.

\bibitem{Jin:1997:WIN}
R.~Jin, M.-E. {Fernandez-Beros}, and R.~P. Novick.
\newblock Why is the initiation nick site of an {AT}-rich rolling circle
  plasmid at the tip of a {GC}-rich cruciform?
\newblock {\em The EMBO Journal}, 16(14):4456--4466, 1997.

\bibitem{KatzS93}
M.~J. Katz and M.~Sharir.
\newblock Optimal slope selection via expanders.
\newblock {\em Information Processing Letters}, 47(3):115--122, 1993.

\bibitem{Kim03}
S.~K. Kim.
\newblock Linear-time algorithm for finding a maximum-density segment of a
  sequence.
\newblock {\em Information Processing Letters}, 86(6):339--342, 2003.

\bibitem{LinHJC03}
Y.-L. Lin, X.~Huang, T.~Jiang, and K.-M. Chao.
\newblock {MAVG}: locating non-overlapping maximum average segments in a given
  sequence.
\newblock {\em Bioinformatics}, 19(1):151--152, 2003.

\bibitem{Lin:2002:EAL}
Y.-L. Lin, T.~Jiang, and K.-M. Chao.
\newblock Algorithms for locating the length-constrained heaviest segments,
  with applications to biomolecular sequence analysis.
\newblock {\em Journal of Computer and System Sciences}, 65(3):570--586, 2002.

\bibitem{Macaya:1976:AOE}
G.~Macaya, J.-P. Thiery, and G.~Bernardi.
\newblock An approach to the organization of eukaryotic genomes at a
  macromolecular level.
\newblock {\em Journal of Molecular Biology}, 108:237--254, 1976.

\bibitem{Madsen:1997:ICE}
C.~S. Madsen, C.~P. Regan, and G.~K. Owens.
\newblock Interaction of {CArG} elements and a {GC}-rich repressor element in
  transcriptional regulation of the smooth muscle myosin heavy chain gene in
  vascular smooth muscle cells.
\newblock {\em Journal of Biological Chemistry}, 272(47):29842--29851, 1997.

\bibitem{Matousek91}
J.~Matou{\v{s}}ek.
\newblock Randomized optimal algorithm for slope selection.
\newblock {\em Information Processing Letters}, 39(4):183--187, 1991.

\bibitem{Murata:2001:TAF}
S.-i. Murata, P.~Herman, and J.~R. Lakowicz.
\newblock Texture analysis of fluorescence lifetime images of {AT}- and
  {GC}-rich regions in nuclei.
\newblock {\em Journal of Hystochemistry and Cytochemistry}, 49:1443--1452,
  2001.

\bibitem{Nekrutenko:2000:ACH}
A.~Nekrutenko and W.-H. Li.
\newblock Assessment of compositional heterogeneity within and between
  eukaryotic genomes.
\newblock {\em Genome Research}, 10:1986--1995, 2000.

\bibitem{OhlerNLR01}
U.~Ohler, H.~Niemann, G.~Liao, and G.~M. Rubin.
\newblock Joint modeling of {DNA} sequence and physical properties to improve
  eukaryotic promoter recognition.
\newblock {\em Bioinformatics}, 17(S1):S199--S206, 2001.

\bibitem{Rice:2000:EEM}
P.~Rice, I.~Longden, and A.~Bleasby.
\newblock {EMBOSS}: The {E}uropean molecular biology open software suite.
\newblock {\em Trends in Genetics}, 16(6):276--277, June 2000.

\bibitem{Scotto:1993:GRD}
L.~Scotto and R.~K. Assoian.
\newblock A {GC}-rich domain with bifunctional effects on {mRNA} and protein
  levels: implications for control of transforming growth factor beta 1
  expression.
\newblock {\em Molecular and Cellular Biology}, 13(6):3588--3597, 1993.

\bibitem{Sellers:1984:PRG}
P.~H. Sellers.
\newblock Pattern recognition in genetic sequences by mismatch density.
\newblock {\em Bulletin of Mathematical Biology}, 46(4):501--514, 1984.

\bibitem{Sharp:1995:DSE}
P.~M. Sharp, M.~Averof, A.~T. Lloyd, G.~Matassi, and J.~F. Peden.
\newblock {DNA} sequence evolution: the sounds of silence.
\newblock {\em Philosophical Transactions of the Royal Society of London Series
  B, Biological Sciences}, 349:241--247, 1995.

\bibitem{Soriano:1983:DIR}
P.~Soriano, M.~Meunier-Rotival, and G.~Bernardi.
\newblock The distribution of interspersed repeats is nonuniform and conserved
  in the mouse and human genomes.
\newblock {\em Proceedings of the National Academy of Sciences of the United
  States of America}, 80:1816--1820, 1983.

\bibitem{Stojanovic:1999:CFM}
N.~Stojanovic, L.~Florea, C.~Riemer, D.~Gumucio, J.~Slightom, M.~Goodman,
  W.~Miller, and R.~Hardison.
\newblock Comparison of five methods for finding conserved sequences in
  multiple alignments of gene regulatory regions.
\newblock {\em Nucleic Acids Research}, 27:3899--3910, 1999.

\bibitem{Sueoka:1988:DMP}
N.~Sueoka.
\newblock Directional mutation pressure and neutral molecular evolution.
\newblock {\em Proceedings of the National Academy of Sciences of the United
  States of America}, 80:1816--1820, 1988.

\bibitem{Wang:2002:RFP}
Z.~Wang, E.~Lazarov, M.~O'Donnel, and M.~F. Goodman.
\newblock Resolving a fidelity paradox: {Why} {Escherichia} coli {DNA}
  polymerase {II} makes more base substitution errors in at- compared to
  {GC}-rich {DNA}.
\newblock {\em Journal of Biological Chemistry}, 277:4446--4454, 2002.

\bibitem{Wolfe:1989:MRD}
K.~H. Wolfe, P.~M. Sharp, and W.-H. Li.
\newblock Mutation rates differ among regions of the mammalian genome.
\newblock {\em Nature}, 337:283--285, 1989.

\bibitem{Wu:1999:IMD}
Y.~Wu, R.~P. Stulp, P.~Elfferich, J.~Osinga, C.~H. Buys, and R.~M. Hofstra.
\newblock Improved mutation detection in {GC}-rich {DNA} fragments by combined
  {DGGE} and {CDGE}.
\newblock {\em Nucleic Acids Research}, 27(15):e9, 1999.

\bibitem{Zoubak:1996:GDH}
S.~Zoubak, O.~Clay, and G.~Bernardi.
\newblock The gene distribution of the human genome.
\newblock {\em Gene}, 174:95--102, 1996.

\end{thebibliography}

\end{document}